\documentclass[12pt]{article}
\usepackage{amsmath}
\usepackage{amssymb}

\makeatletter

\usepackage{amsfonts}

\makeatother
\begin{document}

\title{Dynamical clusters of infinite particle dynamics}

\author{V. A. Malyshev}

\maketitle
\begin{abstract}
For any system $\{ i\}$ of particles with the trajectories $x_{i}(t)$
in $R^{d}$ on a finite time interval $[0,\tau]$ we define the interaction
graph $G$. Vertices of $G$ are the particles, there is an edge between
two particles $i,j$ iff for some $t\in[0,\tau]$ the distance between
particles $i,j$ is not greater than some constant. We undertake a
detailed study of this graph for infinite particle dynamics and prove
exponential estimates for its finite connected components. This solves
continuous percolation problem for a complicated geometrical objects
- the tubes around particle trajectories. 
\end{abstract}

\newpage

\section{\protect\bigskip Introduction}

We undertake a detailed study of the interaction graph for infinite
particle dynamics and prove exponential estimates for its finite clusters.
Cluster properties of infinite-particle dynamics is a folklor notion
now, see \cite{Sin,Lanf,Spohn,CerIllPul}. For classical (deterministic
or stochastic) dynamics, the cluster property reveals a clear geometric
picture, where the system of infinite particles is subdivided into
random finite subsets (clusters) which do not interact between each
other on a fixed time interval. Thus the dynamics is reduced to finite
particle dynamics. The new feature of this paper is that we describe
clusters in detail combinatorially and give combinatorial estimates
of their probabilities. This solves continuous percolation problem
for a complicated geometrical objects - the tubes around particle
trajectories.

Note that cluster property can have also different (but related) meaning:
decay of correlations, existence of quasiparticles, especially for
the quantum case, see \cite{MalMin2}, we do not pursue this issue
here.

We assume that at time $0$, for any finite volume $\Lambda$, Poisson
point field with density $\rho$ is given. Thus the number $N(\Lambda)$
of particles in $\Lambda$ has Poisson distribution with $\frac{<N(\Lambda)>}{\Lambda}=\rho$.
The initial velocities $v_{i}(0)$ of the particles are assumed to
be independently and identically distributed. Random initial configuration
of coordinates $x_{i}(0)$ and velocities $v_{i}(0)$ is denoted by
$\omega=\omega_{\Lambda}$.

We assume that for any cube $\Lambda$ and any fixed number of particles
$N(\Lambda)$ some finite particle dynamics in $\Lambda$ is given.
Here we mean \ by this that for (almost) any initial coordinates
$x_{i}(0)$ and velocities $v_{i}(0)$ the trajectories $x_{i}(t)$
are uniquely defined on the time interval $\left[0,\tau\right]$,
they are assumed to be piecewise smooth (then the velocities $v_{i}(t)=\frac{dx_{i}(t)}{dt}$
are defined a.e.).

We fix further on some $r>0$. We call the tube (or $r$-tube) $T_{i}(\tau)$
of the particle $i$ the $r$-neibourhood of its trajectory $x_{i}(t,\omega),0\leq t\leq\tau$.
We say that two particles $i,j$ interact at time $t$ if \[
dist(x_{i}(t),x_{j}(t))\leq2r\]
 that is if the closed $r$-neighborhoods of $x_{i}(t)$ and $x_{j}(t)$
(the time $t$ slices of the corresponding tubes) intersect.

To define dynamical clusters we consider the following finite random
graph $G^{\Lambda}=G^{\Lambda}(\tau,r)$. Vertices of $G^{\Lambda}=G^{\Lambda}(\tau,r)$
are the particles. We can assume that they are labeled by initial
coordinates of the particles. Two vertices are connected by an edge
if on the time interval $\left[0,\tau\right]$ these two particles
interact at least at one time moment. The connected components of
$G^{\Lambda}=G^{\Lambda}(\tau,r)$ are called dynamical clusters in
$\Lambda$ (or simply clusters) if $\tau$ and $\Lambda$ are fixed.
Equivalently, the connected components are the same as the topological
connected components of the union of all tubes $T=\cup T_{i}(\tau)$
in $\Lambda\times\left[0,\tau\right]$.

Within this general setting, sufficient for our purpose, we have to
do three additional assumptions.

\begin{enumerate}
\item For simplicity of presentation assume that the particle $i$ moves
freely on the time interval $(t,t+s)$\[
x_{i}(t+s]=x_{i}(t)+v_{i}(t+0)s\]
 if for any $t^{\prime}\in(t,t+s)$ this particle does not interact
with other particles. In the absence of interaction one can assume
elastic reflection from the boundary, but this is not necessary as
our estimates will concern large volumes. 
\item For any two particles $i,j$ denote $s_{ij}=s_{ij}(\omega)$ their
first interaction time. We assume that $s_{ij}$ are all different
a.s. This holds trivially in many known dynamical models. 
\item Our main assumption is that the velocities are uniformly bounded,
that is for some constant $v^{0}>0$ and any $i$ and $t\in\left[0,\tau\right]$\[
\left|v_{i}(t)\right|\leq v^{0}\]
 This is very simplifying assumption. However, even under such condition
the combinatorics of clusters is not easy. I hope that the obtained
estimates allow to weaken essentially this condition. 
\end{enumerate}

\paragraph{Physical intuition and percolation theory}

If the velocities are uniformly bounded then on the time interval
$\left[0,\tau\right]$ the tube of a particle belongs to the ball
of radius $v^{0}\tau+r$ with the centre in its initial coordinate.
The results from continuous percolation theory, see \cite{Hall1,Hall2,MeeRoy},
tell us that the infinite system of balls of radius $v^{0}\tau+r$
around Poisson points in $R^{d}$ a.s. has no infinite clusters for
$\rho$ small. More exactly, simple space scaling shows that if $\tau,r$
are fixed, $v^{0}\rightarrow\infty$ and $\rho$ does not exceed $\alpha_{0}(v^{0})^{-d}$
for some fixed factor $\alpha_{0}>0$ sufficiently small, then all
clusters are finite a.s. Moreover, exponential cluster estimates hold.
However this result is too rough for our purpose.

Physics tells us that the correct answer is given by the Boltzman-Grad
(B-G) scaling where the volume $\lambda$ swept up by the particle
is the volume of its tube, that is of order $v^{0}\tau r^{d-1}$,
where $v^{0}\tau$ is the length of the trajectory and $r^{d-1}$
is of the order of circular section of the tube. Heuristically, for
the densities smaller than $\alpha_{0}\lambda^{-1}$, the limiting
dynamical clusters should be finite a.s. The reason is that $N(\Lambda)$
particles sweep up the volume $v^{0}\tau r^{d-1}N(\Lambda)$ which
should be less than $\Lambda$.

As far as I know, this does not follow from the existing results in
continuous percolation theory. One of the reasons is that the particle
tubes are not independent volumes around the Poisson points. We give
the proof here.

\paragraph{Result}

Denote $P_{k}^{\Lambda}(\tau|x)$ the conditional probability that,
under the condition that there is a particle at the point $x$, the
following two conditions hold:

\begin{enumerate}
\item the cluster, containing this particle, has exactly $k$ particles, 
\item these particles do not interact at time $0$, that is the distance
between the particles at time zero is greater than $2d$. 
\end{enumerate}
Normally there can be initial clusters of balls of radius $r$ and
centres $x_{i}(0)$. One of the possibilities to avoid this is to
delete such initial clusters from any configuration. We will pursue
another strategy, which takes these clusters into account. However,
we start with clusters without intersecting balls at time $0$.

Then the following exponential estimate holds.

\begin{description}
\item [Theorem]There are constants $C,\alpha_{0}>0$ such that for any
$\tau,v^{0},r$ and 
\begin{equation}
\rho=\textrm{$\frac{N(\Lambda)}{\Lambda}=\alpha(\tau v^{0}r^{d-1})^{-1}$}
\end{equation}
with
$0<\alpha<\alpha_{0}$ and any $k$ we have, uniformly in $\Lambda$
and in $x$, \begin{equation}
P_{k}^{\Lambda}(\tau|x)<C\alpha^{k-1}\label{main}\end{equation}

\item [Corollary]If for any $t\in[0,\tau]$ the thermodynamic limit of
the dynamics exists, then all clusters in $R^{d}$ are finite and
the exponential estimates (\ref{main}) hold for $P_{k}(\tau)=lim_{\Lambda\rightarrow\infty}P_{k}^{\Lambda}(\tau|x)$. 
\end{description}
Under our general dynamics one cannot prove the existence of thermodynamic
limit (it may not even exist). Under less general conditions it will
be done elsewhere. However, there are many results, obtained in a
different way, concerning the existence of the thermodynamic limit
\cite{Lanf,Spohn,CerIllPul,DobrFri}.

The paper is based on the combinatorial algorithm of {}``construction''
of clusters, using notion of the ordered tree, corresponding to the
cluster. Formal exposition of this idea, presented in the paper, is
not very short, however the reader can get the whole picture by reading
the definitions in the next section and the examples with $N=2,3$.

\section{Combinatorial structure of clusters}

In finite volume $\Lambda$ any cluster is finite, thus for any $x$\[
\sum_{k=1}^{\infty}P_{k}^{\Lambda}(\tau|x)=1\]
 To prove the theorem we estimate the conditional probability (given
$x$) that the initial configuration contains $k$ particles $X=(x_{1}(0),v_{1}(0),...,x_{k}(0),v_{k}(0))$
which would give a cluster with exactly $k$ particles if other particles
in $\Lambda$ were not taken into account. We estimate by $1$ the
conditional probability (given $X$) that there are no other particles
(close to the cluster), which could interact with the particles of
the cluster on the time interval $\left[0,\tau\right]$. We get these
estimates uniformly in $\Lambda$ and $x$.

In this section we describe clusters in combinatorial terms. This
will allow us to estimate the probability of clusters.

\paragraph{Trees of subclusters}

A subgraph of $G^{\Lambda}$ is a subset of vertices together with
all edges between them, inherited from $G^{\Lambda}$. Subcluster
is a connected subgraph of $G^{\Lambda}$. In finite volume any subcluster
is a subgraph of a finite cluster.

Assume that initial coordinates and velocities (that is a point $\omega$
in our probability space) are given in the volume $\Lambda$ with
$N(\Lambda)$ particles. Then the trajectories of particles are uniquely
defined. For any set $A$ of $N=N(A)\leq N(\Lambda)$ particles we
define a tree (of subclusters) $T(A)=T(A,\omega)$ with $2N-1$ vertices.
Assume that initially the distance between any two particles in $A$
is greater than $2d$.

We construct the tree inductively, in $N-1$ steps. To each particle
of $A$ we assign a vertex of lowest level $0$ of the tree $T(A)$,
denote this set of vertices (or one-particle subclusters) $T_{0}$.
Denote $t_{1}=t_{1}(\omega)$ the first moment when a pair of particles
from $A$, denote them $i(1),j(1)$, begin to interact. We say that
these two particles form a subcluster. We associate this subcluster
with the new vertex $w$ of $T(A)$, associate $t_{1}$ with $w$
and connect $w$ with vertices $i(1)$ and $j(1)$ of $T(A)$. Thus
$w$ lies above $i(1)$ and $j(1)$.

Next steps proceed by induction. Assume that $k$ new vertices $w_{1},...,w_{k}$
(subclusters) together with the time sequence\[
t_{1}<...<t_{l}<...<t_{k}\]
 have already been constructed. We shall speak about vertices of the
tree and subclusters interchangebly, the same for particles and vertices
of level $0$.

Let us call maximal subclusters (or maximal vertices of the tree)
after step $k$ the subclusters which are not proper subsets of already
constructed subclusters. Initially there are $N$ (one-particle) maximal
subclusters, and the number of maximal subclusters (maximal vertices)
diminishes \ by $1$ on each step, thus after step $k$ there will
be $N-k$ maximal subclusters. We define $t_{k+1}$ as the next minimal
time moment when the new subcluster $w_{k+1}$ is formed from two
already constructed maximal subclusters denoted $s_{1}(w_{k+1}),s_{2}(w_{k+1})$.
That is two particles $i(k+1)\in s_{1}(w_{k+1}),j(k+1)\in s_{2}(w_{k+1})$
from two different maximal subclusters begin to interact.

This process can stop at some step $k<N-1$, if $A$ is not a subcluster.
Otherwise on the last step $N-1$ we get a subcluster consisting of
$N$ particles, the root vertex of the tree $T(A)$.

Note that all trees $T(A,\omega)$ have the following property. Each
vertex (except those of level $0$) has exactly two adjacent vertices
of lower level. Thus there are exactly $N-1$ vertices of levels different
from $0$, denote the set of these vertices $T_{N-1}$. Denote this
class of trees $\mathcal{T}(N)$. The number of trees in $\mathcal{T}(N)$
does not exceed $C^{N}$ for some absolute constant $C>0$.

Thus, each cluster with $N$ points defines a tree $T\in\mathcal{T}(N)$,
some complete order relation $R$, defined by the time moments $t_{k}$,
on the set $T_{N-1}$, and a function $i(w)$, which assigns to each
vertex $w$ a level $0$ vertex $i(w)$, lying under $w$. The first
interaction of two maximal subclusters $w_{k}$and $w_{l}$ is the
interaction of its particles $i(w_{k})$ and $i(w_{l})$.

Note that we enumerate time moments, subclusters, vertices of the
tree in the order they appear in the inductive process.

Formally we did not assume that the particles from $A$ do not interact
with particles outside $A$ on time interval $[0,\tau]$. However
further we will consider the trees only for this case.

\paragraph{Construction of initial configuration}

To describe all possible initial configurations, which give clusters
with $k$ particles, we proceed in the inverse way. Fix some tree
$T\in\mathcal{T}(N)$ with $N$ vertices of level $0$, one of them
is specified. This vertex will correspond to the particle situated
in $x$. Isomorphism of trees respects the specified vertices. Choice
of the specified particle gives factor $N$ in combinatorial estimates.

Fix also some complete order relation $R$ on $T_{N-1}$, and for
each $w\in T_{N-1}$ a particle $i(w)\in w$. The array\[
\mathbf{B}=(T,R,\{ i(w):w=1,...,N-1\})\]
 will be our underlying combinatorial structure. Now we shall get
an upper bound for the number of such $\mathbf{B}$.

\paragraph{Ordering the tree}

For a given tree $T$ consider the orderings of the tree $T_{N-1}$
(with complete order relation $<$) with the only restriction: $v<v^{\prime}$
if $v$ is under $v^{\prime}$. Let $B(T)\leq(N-1)!$ be the number
of such orderings. Once the tree is ordered, to each vertex of the
ordered tree (except ones of level $0$) we assign some positive number
$0<t_{v}\leq\tau$, satisfying the property: if $v<v^{\prime}$ then
$t_{v}<t_{v^{\prime}}$. The numbers $t_{w}$ are interpreted, as
in the definition of the tree, as first interaction times of two subclusters.

For each time $t_{w}$, corresponding to the interaction of two subclusters
$s_{1}=s_{1}(w)$ and $s_{2}=s_{2}(w)$, one should know also the
pair $i(s_{1}(w))=i(w),j(s_{2}(w))=j(w)$ of particles from these
subclusters which interact at time $t_{w}$. The number of possibilities
gives the factor $D_{s_{1}}D_{s_{2}}$, where $D_{s}$ is the number
of particles in the cluster $s$. In total this gives the factor\[
D(T)=\prod_{w=1}^{N-1}D_{w}\]
 The main combinatorial estimate is the following result, which is
interesting in its own.

\begin{description}
\item [Lemma.]For any tree $T$\begin{equation}
Q(T,N)=B(T)D(T)<C^{N}N!\label{lem1}\end{equation}
 for some absolute constant $C>0$. Moreover, this estimate cannot
be improved.
\end{description}
Proof. We need the following notation. Let $S_{k},S_{l}$ be two completely
ordered sets with $k$ and $l$ elements correspondingly. Denote $R(k,l)$
the number of complete orderings of the set $S_{k}\cup S_{l}$ which
do not change the order inside $S_{k}$ and inside $S_{l}$. Take,
for example, $k\leq l$, then \[
R(k,l)=2\sum_{i=1}^{k}C_{k-1}^{i-1}C_{l-1}^{i-1}\]
 In fact, we can split each of $S_{k}$ and $S_{l}$ on $i=1,...,k$
consecutive nonempty groups and arrange these groups in a sequence
in alternative order. For example, $S_{k}$ can be splitted on $i$
consecutive parts by putting $i-1$ walls on $k-1$ empty places between
consecutive elements of $S_{k}$, that gives the factor $C_{k-1}^{i-1}$.

For any tree $T,\left|T\right|=N$, we have the recurrent relation\begin{equation}
Q(T,N)=NQ(k,T_{1})Q(N-k\,,T_{2})R(k,N-k)\label{rec1}\end{equation}
 if under the root vertex of $T$ there are trees $T_{1}$ and $T_{2}$
with $\left|T_{1}\right|=k,\left|T_{2}\right|=N-k$ correspondingly.
For $q=\log_{2}Q$ we have\begin{equation}
q(T,N)=\log N+q(k,T_{1})+q(N-k\,,T_{2})+\log R(k,N-k)\label{rec2}\end{equation}

One can easily get uniform estimates separately for $B(T)\leq N!$
and $D(T)\leq N!$, but it is too rough, because there are cases where
$B(T)=N!,D(T)=C^{N}$ and vice versa. This is seen from the following
two examples. In the first one for the sequence of subclusters\[
\left\{ 1,2\right\} ,\left\{ 1,2,3\right\} ,....,\left\{ 1,2,...,N-1\right\} \]
 we have $B(T)=1,D(T)=(N-1)!$.

From the second example on sees, moreover, that the estimate (\ref{lem1})
cannot be improved. Put $N=2^{n}$ and consider the tree $T$ with
$2^{n-k}$ vertices on levels $k=0,...,n$. Denote $Q(n)=B(T)D(T)$.
We have the recurrent relation\[
Q(n)=2^{2n}Q^{2}(n-1)R(2^{n-1},2^{n-1})\]
 It follows\[
Q(n)\leq2^{2n}Q^{2}(n-1)a\frac{2^{N}}{\sqrt{N}}\]
 for some constant $a>0$. For $q(n)=\log_{2}Q(n)$ this gives for
$n\geq2$\[
q(n)\leq2q(n-1)+2^{n}+bn,q(1)=1\]
 for some constant $b>0$. The solution of this inequality is\[
q(n)\leq\sum_{k=1}^{n}2^{k}(2^{n-k}+b(n-k))=n2^{n}+b2^{n}\sum_{k=1}^{n}(n-k)2^{-n+k}=(n+c)2^{n}\]
 for some $c>0$. This gives\[
Q(N)\leq2^{NlogN+cN}\]

Now we come to the general case. Put $Q(N)=P(N)N!$, then from (\ref{rec1})
we get\[
P(N)\leq NP(k)P(N-k)r(k,N-k)\]
 where\[
r(k,N-k)=\frac{2\sum_{i=1}^{k}C_{k-1}^{i-1}C_{N-k-1}^{i-1}}{C_{N}^{k}}\]
 for some $k\leq[\frac{N}{2}]$. For $p(N)=logP(N)$ we have\[
p(N)=p(k)+p(N-k)+a(v)\]
 \[
a(v)=a(k,N-k)=log(Nr(k,N-k))\]
 This equation can be solved explicitely as\[
p(N)=\Sigma_{v}a(v)\]

To make estimation of this sum we need some notation and results.
In the inductive procedure for a given tree $T$ we will distinguish
vertices of type $A$ or $B$, where correspondingly $k\in\left[\alpha\left[\frac{N}{2}\right],\left[\frac{N}{2}\right]\right]$
and $k\in\left[0,\alpha\left[\frac{N}{2}\right]\right]$ where $\alpha=1-\varepsilon$
for some small $\varepsilon>0$. Denote their numbers $N_{A},N_{B}$
correspondingly. We have $N_{A}+N_{B}=N-1$.

Introduce the depth $m(v)$ of the vertex $v$ of the tree - the distance
from the root vertex. It is clear that $\log N\leq m(v)\leq N$. The
$A$-depth $m_{A}(v)$ of the $A$-vertex $v$ is the number of $A$-vertices
on the path from it to the root. We have\begin{equation}
m_{A}(v)<log_{b}N,b=\frac{1}{2}(1+\epsilon)\label{A1}\end{equation}
 The number\begin{equation}
\#(v:m_{A}(v)=m)<c^{m},c=2(1+\epsilon)\label{A2}\end{equation}
 Denote $N(v)$ the number of level $0$ vertices under $v$. We have\begin{equation}
c_{-}^{m(v)}<N(v)<Nc_{+}^{m(v)}\label{A3}\end{equation}
 where\[
c_{-}=\frac{1}{2}(1-\epsilon),c_{+}=\frac{1}{2}(1+\epsilon)\]

We will need some inequalities.

\begin{enumerate}
\item For any $k,N$ \begin{equation}
\frac{2\sum_{i=1}^{k}C_{k-1}^{i-1}C_{N-k-1}^{i-1}}{C_{N}^{k}}\leq1\label{B1}\end{equation}
 This can be proved by simple combinatorial argument. Take four intervals
$1,2,[3,k+1],[k+2;N]$. Then we choose $k$ elements from these two
intervals. The number $2C_{k-1}^{i-1}C_{N-k-1}^{i-1}$ gives only
restricted choice: we choose one of the first two elements (factor
$2$), $i-1$ elements from the last interval (factor $C_{N-k-1}^{i-1}$)
and $k-i$ elements from the third interval (factor $C_{k-1}^{i-1}$). 
\item (large deviation estimate) We will use the asymptotics\[
\log C_{N}^{\alpha N}\sim H(\alpha)N,H(\alpha)=-\alpha\log\alpha-(1-\alpha)\log(1-\alpha)\]
 Then for $\gamma\leq\beta\leq\frac{1}{2}$ the maximum of\[
\log C_{\beta N}^{\gamma N}+\log C_{(1-\beta)N}^{\gamma N}\]
 is attained for $\gamma$ satisfying\[
\frac{1}{\beta}log\frac{\beta-\gamma}{\gamma}+\frac{1}{1-\beta}log\frac{1-\beta-\gamma}{\gamma}=0\]
 In fact,\[
\log C_{\beta N}^{\gamma N}+\log C_{(1-\beta)N}^{\gamma N}\sim N(H(\frac{\gamma}{\beta})\beta+H(\frac{\gamma}{1-\beta})(1-\beta))\]
 \[
\frac{d}{d\gamma}(H(\frac{\gamma}{\beta})\beta+H(\frac{\gamma}{1-\beta})(1-\beta))=\frac{1}{\beta}log\frac{\beta-\gamma}{\gamma}+\frac{1}{1-\beta}log\frac{1-\beta-\gamma}{\gamma}\]
 Then by large deviation principle for any $k\sim\beta N,\beta<\frac{1}{2}$\begin{equation}
\log r(k,N-k)<(1-\delta)^{N}\label{A4}\end{equation}

\end{enumerate}
We estimate $a_{v}<N$ for $B$-vertices and $a_{v}<N(v)c_{+}^{m(v)}$
for $A$-vertices. Using the bounds \ref{A1},\ref{A2},\ref{A3},\ref{A4},\ref{B1}
we get the proof.

\section{Probability of clusters}

For any $\mathbf{B}$ and any sequence $0<t_{1}<...<t_{N-1}\leq\tau$
we define the set $I(\mathbf{B},t_{1},...,t_{N-1})$ of initial configurations
$\omega$ of particles such that there exists a set $A,\left|A\right|=N$,
among them (containing some particle at $x$) such that $A$ defines
a cluster, $N(A)=N$ and $T(A,\omega)=T$. The order of $t_{k}$,
of course, should be compatible with the order $R$.

We use the estimate\[
P_{N}^{\Lambda}(\tau|x)\leq\sum_{\mathbf{B}}P(\cup_{t_{1},...,t_{N-1}}I(\mathbf{B},t_{1},...,t_{N-1}|x))\]
 In the previous section we have proved the factorial estimate of
the number of terms in the sum $\sum_{\mathbf{B}}$. Now we will get
the bound uniform in $\mathbf{B}$\[
P(\cup_{t_{1},...,t_{N-1}}I(\mathbf{B},t_{1},...,t_{N-1})|x)\leq C\frac{\alpha^{N-1}}{N!}\]

To estimate the probability $P(\cup_{t_{1},...,t_{N-1}}I(\mathbf{B},t_{1},...,t_{N-1})|x)$
we will find the set\[
J(\mathbf{B)=}\cup_{t_{1},...,t_{N-1}}J(\mathbf{B},t_{1},...,t_{N-1})|x)\subset\Lambda^{k-1}\]
 (further on one of the coordinates is fixed to be $x)$ of all possible
initial configurations of $N=\left|A\right|$ particles which form
the cluster, other particles will be ignored, that is we estimate
the corresponding conditional probabilities by $1$. To estimate the
volume of $J(\mathbf{B}))$, we use the fact that the probability
measure on $\Lambda^{k-1}$ is a product of (Poisson) measures on
$\Lambda$.

We shall see that the measure on the simplex $\left\{ (t_{1},...,t_{N-1}):\sum t_{i}=\tau\right\} $given
by\[
J(A)=\cup_{\left\{ t_{1},...,t_{N-1}\right\} \in A}I(\mathbf{B},t_{1},...,t_{N-1}|x)\]
 is absolutely continuous with respect to Lebesgue measure on the
simplex. By estimating marginal measures we will get the desired bound
for the volume.

We shall explain this first for $N=2,3$.

\paragraph{Two particle cluster}

For $N=1$ the cluster is a tube, corresponding to the trajectory
$x_{1}(t)=x+v_{1}t$. We estimate the conditional probability, that
no other particle interacts with this one, by $1$.

Consider now two particles $1,2$, with initial velocities $v_{1},v_{2}$.
Assume that the initial coordinate of the particle $1$ is $x_{1}(0)=x$.
Denote $S_{r}(x)$ the $(d-1)$-dimensional sphere of radius $r$
and centre $x$, let $B_{r}(x)$ be the corresponding $d$-dimensional
ball. Put $S_{r}=S_{r}(0),B_{r}=B_{r}(0)$. If $t_{1}$ (the first
contact time) is given then the relative initial coordinate $x_{2}(0)-x_{1}(0)$
of the particle $2$ can take any value\[
x_{2}(0)-x_{1}(0)=y+x_{1}(t_{1})-x_{2}(t_{1})=y+x+t_{1}(v_{1}-v_{2})\]
 where $y$ is any point on the sphere $S_{2r}$ such that for any
$0\leq t<t_{1}$ the spheres $S_{r}(x_{1}(t))$ and $S_{r}(x_{2}(t))$
do not intersect.

Thus $x_{2}(0)$, for given $v_{1},v_{2},t_{1},x_{1}(0)$, should
be on a subset of some sphere\[
S_{2r}(x+t_{1}(v_{1}-v_{2}))=\left\{ y+x+t_{1}(v_{1}-v_{2}):y\in S_{2r}\right\} \]
 in $R^{d}$. The union \[
\cup_{t_{1}\in\left[0,\tau\right]}S(x+t_{1}(v_{1}-v_{2}))\]
 is a closed subset of $R^{d}$ with some volume $V$. This volume,
uniformly in $v_{1},v_{2}$, does not exceed $\beta\tau$ where $\beta=\beta(d,r,v^{0})=C_{d}v^{0}r^{d-1}$,
where $C_{d}$ is some absolute constant. That is obtained by integration
in $t_{1}$. Thus the probability that initially there is a particle
in this volume does not exceeed\[
1-\exp(-C_{d}\alpha)\]
 for $c=c(d)>0$. This gives $O(\alpha)$ if $\alpha$ is small. Given
these two particles, we estimate from above the conditional probability
that there are no more particles in their vicinity so that they either
could form initial cluster or could interact with them on the time
interval $(0,\tau]$, by $1$.

\paragraph{Formation of the second subcluster}

Assume now that there are more than $2$ particles. Assume that the
formation of the first subcluster occurs at time $t_{1}$, we enumerate
the two particles as $1,2$. Note that, by definition of the tree,
all particles besides $1,2$ will move freely until the time $t_{2}$,
and the dynamics of the pair in the first subcluster is independent
of the other particles until the time moment $t_{2}$. At this moment
there can be two cases: two new particles $3,4$ will form subcluster
at $t_{2}$, or subcluster $1,2,3$ will be formed at time $t_{2}$.
Note however, that the case is fixed if the tree is given.

Consider the volume $J(\mathbf{B})$. We will see that in both cases,
for fixed $t_{1}$ the volume of the $t_{1}$-slice of $J(\mathbf{B})$
does not exceed $\beta(\tau-t_{1})$. It follows that the volume $J(\mathbf{B})$
cannot exceed $\beta Q_{2}$, where $Q_{2}$ is the volume of the
simplex $t_{1}+t_{2}=\tau$.

In the first case we fix the relative coordinate $x_{3}(0)-x_{4}(0)$
as previously. However, we do not get a cluster.

In the second case let the interacting particles at time $t_{2}$
be $1$ and $3$. Let us assume that $x_{1}(0)=x$, because we can
make a global translation of the vector\[
(x_{1}(0),...,x_{N}(0))\rightarrow(x_{1}(0)-x_{i}(0)+x,...,x_{N}(0)-x_{i}(0)+x)\]
 afterwards, thus fixing$x\textrm{$_{\textrm{i}}(0)=x$}$. Thus we
know $x_{1}(t_{1})$ and $x_{1}(t_{2})$, and we should find the volume
$V_{3}(0)$ where $x_{3}(0)$ should be situated so that the interaction
could occur. It is sufficient to find the volume $V_{3}(t_{1})$ where
$x_{3}(t_{1})$ can be because $V_{3}(0)=V_{3}(t_{1})-v_{3}t_{1}$.
But $V_{3}(t_{1})$ is the union of subsets of the shifted spheres\[
V_{3}(t_{1})=\cup_{t_{2}\in(t_{1},\tau)}(S_{2r}+(t_{2}-t_{1})v_{3})\]
 Integration in $t_{2}$ gives the desired estimate.

\paragraph{General case}

Afterwards we proceed by induction. During this procedure we will
fix relative coordinates inside the subclusters. Assume that the corresponding
subclusters have already been constructed for all times $t_{1},...,t_{k}$.
The relative coordinates inside each maximal cluster have already
been fixed. There will be $m\leq N-k$ maximal subclusters. In each
maximal subcluster $s$ one particle is specified $i(s)$. The partilce
$i(s)$ is one which will interact at the first time in this maximal
cluster with a particles ourside this subcluster.

On the step $k+1$ we assume that clusters $w$ and $z$ interact,
merely the particles $a(w)$ and $a(z)$ of these clusters correspondingly.
will fix relative coordinate $x_{a(w)}(t_{k})-x_{a(z)}(t_{k})$. Note
that $X(w,t_{k+1})=x_{a(w)}(t_{k+1})-x_{a(w)}(t_{k})$ does not depend
on $x_{a(w)}(t_{k})$ if the relative coordinates of the particles
inside $w$ are fixed. Similarly for $X(z,t_{k+1})=x_{a(z)}(t_{k+1})-x_{a(z)}(t_{k})$.
Then \[
x_{a(w)}(t_{k})-x_{a(z)}(t_{k})=y+X(z,t_{k+1})-X(w,t_{k+1})\]
 where $y$ is any point on $S_{2r}$ such that for any $t_{k}\leq t<t_{k+1}$
the spheres $S_{r}(x_{a_{i}}(t))$ and $S_{r}(x_{a_{j}}(t))$ do not
intersect.

We want to prove as earlier that for fixed $t_{1},...,t_{k-1},t_{k+1},...,t_{N-1}$
the subset on the $t_{k}$-axis has measure not exceeding $c(t_{k+1}-t_{k-1})$.
From this it follows that the volume is of order of the volume of
the simplex itself and thus is equal to\[
I(N-1)=\int_{0}^{\tau}...\int_{t_{N-3}}^{\tau}\int_{t_{N-2}}^{\tau}dt_{N-1}...dt_{1}=\int_{0}^{\tau}...\int_{0}^{t_{3}}\int_{0}^{t_{2}}dt_{1}...dt_{N-1}=\frac{\tau^{N-2}}{(N-2)!}\]

\paragraph{Cases with initial clusters}

Assume now that initially there are particles with the distance less
or equal $2d$. Consider the graph $G^{\Lambda}(0)$, its connected
components are called initial subclusters. We assign with them $M<N$
vertices and proceed as before, thus we get time moments\[
t_{1}<...<t_{M-1}\]

\section{Example of dynamics}

Here we give an examples of dynamics (with chemical reactions), satisfying
the above general conditions. This will be however random dynamics.
In this dynamics randomness occurs because of random initial conditions
and random interaction rules of dynamics. Denote the corresponding
random elements of the probability space $\omega$ and $\omega_{1}(\omega)$.
The above results will hold uniformly in $\omega_{1}$.

Define first the finite volume dynamics. There are $N<\infty$ particles,
each particle is characterized at time $t\in R_{+}$ by its coordinate
$x(t)\in\Lambda\subset R^{d}$, velocity $v(t)\in R^{d}$ and type
$a(t)\in\left\{ 1,...,A\right\} $. Initial coordinates $x_{i}(0)$
are distributed uniformly in $\Lambda$. The vectors $(a_{i}(0),v_{i}(0))$
are independently distributed with densities $p_{a}(v)$\[
\sum_{a}\int p_{a}(v)dv=1\]
 Any pair of particles $i,j$ at any time interval $(t,t+dt)$ can
change their types and velocities with rates\[
\lambda(Y_{i}(t),Y_{j}(t))\chi_{2r}(x_{i}(t)-x_{j}(t))\]
 where $Y_{i}(t)=(x_{i}(t),a_{i}(t).v_{i}(t)),\chi_{2r}(x)=1$ if
$\left|x\right|\leq2r$, and $0$ otherwise. Functions $\lambda$
are assumed to be bounded. As a result of this jump coordinates do
not change, but types and velocities change\[
a_{i},v_{i},a_{j},v_{j}\rightarrow a_{i}^{\prime},v_{i}^{\prime},a_{j}^{\prime},v_{j}^{\prime}\]
 via conditional probability densities $P(a_{i}^{\prime},v_{i}^{\prime},a_{j}^{\prime},v_{j}^{\prime}|Y_{i},Y_{j})$
so that for any $Y_{i},Y_{j}$ we have\[
\sum_{a_{i}^{\prime},a_{j}^{\prime}}\int P(a_{i}^{\prime},v_{i}^{\prime},a_{j}^{\prime},v_{j}^{\prime}|a_{i},v_{i},a_{j},v_{j})dv_{i}^{\prime}dv_{j}^{\prime}=1\]
 Our main assumption that velocities are uniformly bounded that is\[
P(a_{i}^{\prime},v_{i}^{\prime},a_{j}^{\prime},v_{j}^{\prime}|a_{i},v_{i},a_{j},v_{j})=0\]
 if $v_{i}^{\prime}$ or $v_{j}^{\prime}$ exceed some constant $V_{0}$.
This can be interpreted as redistribution of kinetic enrgy and enrgy
of some chemical bonds of the particles occurs so that kinetic energy
remains bounded.

Velocities and types are piecewise constant on $\left[0,\infty\right]$,
jumps occur at discrete time moments \[
0<t_{1}(\omega)<...<t_{k}(\omega)<...\]
 The following continuous time Markov process $(x_{i}(t),v_{i}(t),a_{i}(t):i=1,2,...,N)$
together with initial data $(x_{i}(0),v_{i}(0),a_{i}(0):i=1,2,...,N)$
is defined above together with\[
x_{i}(t,\omega)=x_{i}(0,\omega)+\int_{0}^{t}v_{i}(t,\omega)dt\]
 The process is well-defined - $x_{i}(t,\omega)$ exist for any initial
data and are piecewise linear a.s. This means that between the jumps
the particles move freely with constant velocities. Then the velocities
$v_{i}(t)=\frac{dx_{i}(t)}{dt}$ are defined a.e.

\end{document}